\documentclass{bconf99}

\newcommand{\klgg}    {\mbox{$K^\circ_L \! \rightarrow \!  \gamma\gamma$ }}
\newcommand{\klmm}    {\mbox{$K^\circ_L \! \rightarrow \! \mu^+ \mu^-$ }}
\newcommand{\klee}    {\mbox{$K^\circ_L \! \rightarrow \! e^+ e^-$ }}
\newcommand{\klme}    {\mbox{$K^\circ_L \! \rightarrow \! \mu e$}}
\newcommand{\kpme}    {\mbox{$K^+ \! \rightarrow \! \pi^+ \mu^+ e^-$}}
\newcommand{\klpme}   {\mbox{$K^\circ_L \! \rightarrow \! \pi^\circ \mu e$ }}
\newcommand{\klpnn}   {\mbox{$K^\circ_L \! \rightarrow \! \pi^\circ \nu \overline{\nu}$ }}
\newcommand{\kzpnn}    {\mbox{$K \! \rightarrow \! \pi \nu \overline{\nu}$ }}
\newcommand{\klpll}   {\mbox{$K^\circ_L \! \rightarrow \! \pi^\circ \ell^+ \ell^-$ }}
\newcommand{\klpee}   {\mbox{$K^\circ_L \! \rightarrow \! \pi^\circ e^+ e^-$ }}
\newcommand{\klpmm}   {\mbox{$K^\circ_L \! \rightarrow \! \pi^\circ \mu^+ \mu^-$ }}
\newcommand{\kpnn}    {\mbox{$K^+ \! \rightarrow \! \pi^+ \nu \overline{\nu}$ }}
\newcommand{\kleeg}   {\mbox{$K^\circ_L \! \rightarrow \! e^+ e^-\gamma$ }}
\newcommand{\klmmee}  {\mbox{$K^\circ_L \! \rightarrow \! \mu^+ \mu^- e^+ e^-$ }}
\newcommand{\klllg}   {\mbox{$K^\circ_L \! \rightarrow \! \ell^+ \ell^- \gamma$ }}
\newcommand{\klllll}  {\mbox{$K^\circ_L \! \rightarrow \! \ell^+ \ell^- \ell'^+ \ell'^-$ }}
\newcommand{\bpsiks}  {\mbox{$B^\circ_d \! \rightarrow \! \psi K^\circ_S$ }}
\newcommand{\kpen}    {\mbox{$K^+\!\rightarrow\!\pi^\circ e^+\nu_e$ }}
\newcommand{\kpp}     {\mbox{$K^+ \! \rightarrow \! \pi^+ \pi^\circ$ }}
\newcommand{\klpen}   {\mbox{$K^\circ_L \! \rightarrow \! \pi^\pm e^\mp \nu_e$ }}
\newcommand{\vtd}     {\mbox{$V_{td}$}}

\newcommand{\vcd}     {\mbox{$V_{cd}$ }}

\begin{document}

\title{Rare Kaon Decays
\footnote{ To be published in the Proceedings of The 3$^{rd}$ International Conference on 
B Physics and CP Violation, Taipei, Taiwan, December 3--7, 1999, H. -Y. Cheng and 
W. -S. Hou, eds. (World Scientific, 2000).}
}

\author{S. Kettell}

\address{Brookhaven National Laboratory\\ Upton, NY 11973-5000}

\maketitle

\abstracts{ The current status of rare kaon decay experiments is
reviewed.  New limits in the search for Lepton Flavor Violation are
discussed, as are new measurements of the CKM matrix.}

\section{Introduction}

The field of rare kaon decays has a long and rich history.  As the
field has evolved so has the definition of `rare'. In this review 
it refers to
branching ratios less than $\sim10^{-10}$. The two areas of 
greatest interest have been the very sensitive searches for physics
beyond the Standard Model (SM) through lepton flavor violating (LFV)
decays and the studies of the SM picture of CKM mixing and CP
violation that have recently begun to come to fruition.

\section{Lepton Flavor Violating Decays}

There is solid experimental evidence for an additive quantum
number associated with each family of charged leptons.  While 
there is no SM mechanism for LFV (even if
m$_{\nu}$$\ne$0, LFV is too small to observe in the charged lepton
sector), there is no underlying gauge symmetry preserving lepton
flavor.  Observation of LFV would be unambiguous evidence for physics
beyond the SM, and is predicted by many extensions to the SM.

Due to excellent sensitivity, the mass scale probed by rare kaon decay
experiments is quite large.  This can be seen by comparing the \klme\
decay, through a hypothetical LFV vector boson with coupling $g_X$ and mass
$M_X$ to the conventional
$K_{\mu2}$ decay ($g$ and $M_W$); a lower limit on $M_X$ can be derived:
\begin{eqnarray}
M_X & \simeq & \frac{g_X}{g} (\sin^2\theta_c)^{1/4} M_W 
\left[ \frac{\Gamma(K^+\rightarrow\mu^+\nu)}{\Gamma(\klme)}\right]^{1/4} \\ \nonumber
 & \simeq & 200 {\rm TeV/c^2} \times \frac{g_X}{g} \times 
\left[ \frac{10^{-12}}{\rm B(\klme)}\right]^{1/4}
\label{eq:lfv}
\end{eqnarray}

\boldmath
\subsection{\klme}
\unboldmath

\begin{figure}[ht]
\epsfxsize=4.5in 
\epsfysize=2.5in 
\epsfbox{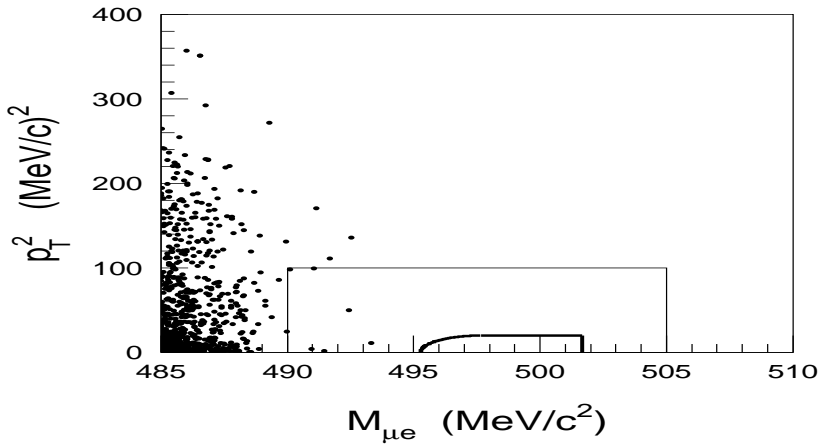} 
\caption{Final E871 data sample after all cuts, with no events in the
signal region. The exclusion box was used to set cuts in an unbiased
way on data far from the signal region. The shape of the signal box
was optimized to maximize signal/background.
\label{fig:e871}}
\end{figure}

The E871 experiment at BNL has finished, with two long runs in
1995--96. No events were seen in the signal region, with an expected
background of 0.1 events (see Fig.~\ref{fig:e871}),
and the 90\% CL limit\cite{e871_me} is B(\klme) $< 4.7\times10^{-12}$.  There
are no plans to pursue this decay further; it is expected
to be limited by background from \klpen with a $\pi\rightarrow\mu$
decay and an electron scattering in the vacuum window or first
tracking chamber at $\sim10^{-13}$.

\boldmath
\subsection{\kpme}
\unboldmath

\begin{figure}[ht]
\vspace{-0.7cm}
\epsfxsize=4.75in 
\epsfysize=3.in
\epsfbox{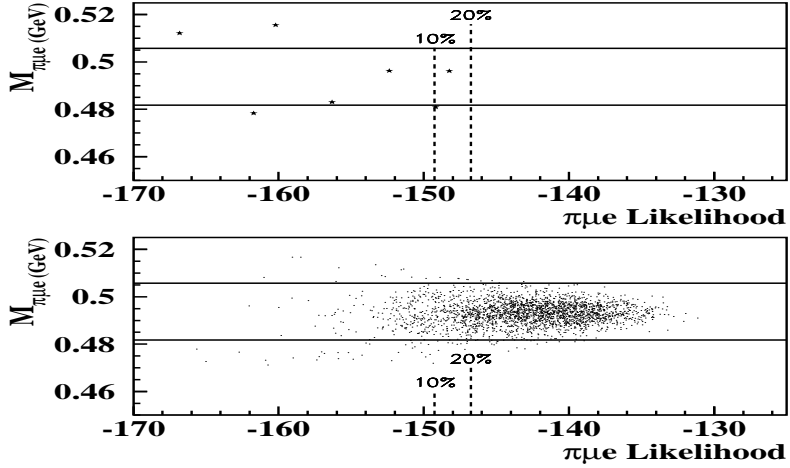} 
\vspace{-1.cm}
\caption{Final 1996 data sample after all cuts, with no events above
$\pi\mu e$ likelihood of 20\%. Also shown are $\pi\mu e$ Monte Carlo
events passing all cuts.  \label{fig:e865}}
\end{figure}

The E865 experiment at BNL has collected data for the decay \kpme\
during the 1995, 1996, and 1998 runs of the AGS. The 90\% CL limit on this
mode from the 1995 run\cite{e865_95}, similar in sensitivity to the predecessor
experiment E777, was B(\kpme) $< 2\times10^{-10}$.  From the
1996 run\cite{e865_pme} with no events above a $\pi\mu e$ likelihood
$>$ 20\%, the limit is B(\kpme) $< 4.8\times10^{-11}$ (see Fig.~\ref{fig:e865}).
Combining the results from E777 and the E865 runs in 1995 and 1996 a
limit of B(\kpme) $ < 3.2\times10^{-11}$ is obtained.  The final
sensitivity, including 1998 data, is expected to be $\sim$3
times better.  There are no plans to continue with this search.

\boldmath
\subsection{\klpme}
\unboldmath

The current limit on \klpme from E799-I\cite{e799_pme} at FNAL, is
B(\klpme) $< 3.1\times10^{-9}$ (90\% CL). This measurement has very small
background levels, so E799-II (KTeV) will be able to substantially
improve upon this limit.

\boldmath
\section{\klmm and other Semi-Rare Decays}
\unboldmath

Two modes for which it may be possible to extract fundamental
CKM parameters, from measurement of the short
distance contributions, are \klmm and \klpll. The \klpll modes are
covered in detail by Taku Yamanaka elsewhere in this volume.  The
current 90\% CL limit\cite{e799_pee} for \klpee is B(\klpee) $ <
5.6\times10^{-10}$ and the limit\cite{e799_pmm} for \klpmm is
B(\klpmm) $ < 3.4\times10^{-10}$.  Both of these measurements have
background which will slow improvement in these modes.

\boldmath
\subsection{\klmm}
\unboldmath

The mode $K^\circ_L \! \rightarrow \! \mu^+ \mu^-$, which has played
such an important role in the development of the SM (e.g. the GIM
mechanism and the prediction of the charm quark), has now been
measured to the unprecedented precision of 1.5\%, with 6200 events, by
E871.\cite{e871_mm}

The decay \klmm is dominated by \klgg with the two real photons converting
to a $\mu^+$$\mu^-$ pair. This contribution is precisely calculated
using QED from a measurement of the \klgg branching ratio. There is
also a long distance dispersive contribution, through off-shell
photons. Most interesting is the short distance contribution, through
internal quark loops, dominated by the top quark.  A measurement of
this short distance contribution is sensitive to the real part of the
elusive \vtd, and will determine $\rho$:
\begin{equation}
B_{SD}(\klmm) = 1.51 \times 10^{-9} A^4 (\rho_0-\overline{\rho})^2
\end{equation}
with $\rho_0 = 1.2$, $\overline{\rho} = \rho(1-\lambda^2/2)$ and
$\lambda$, A, $\rho$ from the Wolfenstein parameterization of the CKM
matrix\cite{wolfenstein}. The current measurement of the branching
ratio B(\klmm) $= (7.18 \pm 0.17) \times 10^{-9}$ by the E871
collaboration\cite{e871_mm} represents a factor of three improvement
in the uncertainty and the error on B(\klmm) no longer dominates the
error on the ratio:
\begin{eqnarray}
\frac{\Gamma(K^\circ_L \! \rightarrow \! \mu\mu)}
{\Gamma(K^\circ_L \! \rightarrow \!  \gamma\gamma)} & = &
\begin{array}{c}B(K^\circ_L \! \rightarrow \! \mu^+ \mu^-) \\ 
\hline B(K^\circ_L \! \rightarrow \! \pi^+ \pi^-)\end{array} 
 \left| \begin{array}{c}\eta_{+-}\\ \hline \eta_{\circ\circ} \end{array} \right|
 \begin{array}{c}B(K^\circ_S \! \rightarrow \! \pi^+ \pi^-)
\\ \hline B(K^\circ_{\rm S} \! \rightarrow \! \pi^\circ \pi^\circ) \end{array} 
 \begin{array}{c}B(K^\circ_{\rm L} \! \rightarrow \! \pi^\circ \pi^\circ)
\\ \hline B(K^\circ_L \! \rightarrow \!  \gamma\gamma)
\end{array}  \\ \nonumber
 & = &  [(3.474\pm.054)\times10^{-6}][1.004\pm.002]
[2.19\pm.03][.632\pm.009] \\ \nonumber
& & [1.55\%] [0.23\%][1.28\%][1.42\%] \\ \nonumber
& = & (1.213\pm0.030)\times10^{-5} \nonumber
\label{eqn_kmm}
\end{eqnarray}
This value is only slightly above the unitarity bound from the
on-shell two photon contribution
\begin{equation}
\frac{\Gamma(K^\circ_L \! \rightarrow \! \mu^+ \mu^-)}
{\Gamma(K^\circ_L \! \rightarrow \!  \gamma\gamma)} > 1.195 \times 10^{-5}
\end{equation}
and leaves very little room for the short distance contribution.
Currently, with conservative estimates of the long distance dispersive
contribution, a limit on $\rho$ can be extracted $\rho >-0.33$ 
(90\% CL)\cite{e871_mm}.

Unlike \klmm, the decay \klee is predominantly through two off-shell
photons, so this decay is less interesting for
extracting SM parameters. However, the recent observation of four
events by E871\cite{e871_ee}, with B(\klee) $= (8.7^{+5.7}_{-4.1})
\times 10^{-12}$ is consistent with Chiral Perturbation Theory
(ChPT) predictions\cite{valencia,dumm} and is the smallest branching
ratio ever measured.

\boldmath
\subsection{\kleeg, \klmmee} 
\unboldmath

Many new measurements of radiative kaon decays have been reported
recently.\cite{daphne}  In particular, new measurements of \klllg and
\klllll have been made by NA48 and KTeV. These measurements are
substantially improved over previous values and even larger
improvements will be obtained when the complete data sets are
analyzed. For example, the new measurement from NA48\cite{na48_eeg}
B(\kleeg) $ = (1.06\pm0.02\pm0.02\pm0.04)\!\times\!10^{-5}$ is
$\sim$70 times more sensitive than the previous measurement and KTeV
will improve this by an additional factor of $\sim$20 beyond NA48.  These
modes, which are not rare, with branching ratios from
$10^{-8}$--$10^{-5}$, may help to determine the long distance
dispersive contribution to \klmm. Additional work in ChPT is needed to
extract the short distance physics.\cite{valencia,dambrosio}

\boldmath
\section{\kzpnn}
\unboldmath

The \kzpnn modes --- \kpnn and \klpnn --- are the `golden modes' for
determining the CKM parameters $\rho$ and $\eta$; and, along with the
other golden mode \bpsiks and perhaps
$\Delta_{m_{B_s}}/\Delta_{m_{B_d}}$, provide the best opportunity
to over-constrain the unitary triangle and to search for new physics.
The unitarity of the CKM matrix can be expressed as
\begin{equation}
V^*_{us}V_{ud} + V^*_{cs}V_{cd} + V^*_{ts}V_{td} = 0
\end{equation}
with the three vectors $V^*_{is}V_{id}$ converging to
form an elongated triangle in the complex plane. The first vector
$V^*_{us}V_{ud}$ is well determined from the decay \kpen.
The height  will be measured by \klpnn and the
third side $V^*_{ts}V_{td}$ will be measured by \kpnn.
The \kzpnn decays are sensitive to the magnitude and imaginary part of \vtd. 
From these two modes the unitarity triangle can be completely
determined.

The theoretical uncertainty in \kpnn ($\sim$7\%) is small and even
\nolinebreak[4]{smaller} in \klpnn ($\sim$1\%); the hadronic matrix
element can be extracted from the well measured \kpen ($K_{e3}$)
decay.  The branching ratios have been calculated to
next-to-leading-log approximation,\cite{bb1} complete with isospin
violation corrections\cite{marciano} and two-loop-electroweak
effects.\cite{bb2} Based on current understandings of SM parameters,
the branching ratios can be expressed as:\cite{bb3}
\begin{eqnarray}
B(\kpnn) & = &\frac{\kappa_+ \alpha^2\!B(K_{e3})}
        {2\!\pi^2\!\sin^4\!\theta_W |V_{us}|^2}\!
      \sum_l  |X_t^l(x_t)V^*_{ts}V_{td}\!+\!X_c^l(x_c)V^*_{cs}\vcd\!|^2\\ \nonumber
         & = & 8.88 \times 10^{-11} A^4 
[(\overline{\rho}_0-\overline{\rho})^2 + (\sigma\overline{\eta})^2] \\ \nonumber
         & = & (0.82\pm0.32)\times10^{-10} \\
B(\klpnn) & = & \frac{\tau_{K_L}}{\tau_{K^+}}
\frac{\kappa_L\alpha^2 B(K_{e3})}{2\pi^2sin^4\theta_W |V_{us}|^2}
\sum_{l} |Im(V^*_{ts}V_{td})X(x_t)|^2 \\ \nonumber
 & = & 4.08\times10^{-10}A^4\eta^2 \\ \nonumber
 & = & (3.1\pm1.3)\times10^{-11} 
\end{eqnarray}
(see Ref~\ref{ref:bb3} for definitions).
The decay \klpnn is direct CP violating, and offers the best opportunity
for measuring the Jarlskog invariant $J_{CP}$,\cite{jarlskog} the
fundamental measure of CP violation in the SM.

\boldmath
\subsection{\kpnn}
\unboldmath

\begin{figure}[ht]
\hspace{1.cm}
\epsfxsize=4.in
\epsfysize=2.25in 
\epsfbox{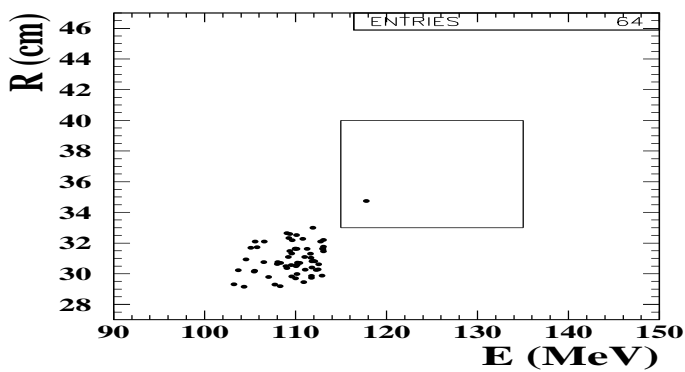}
\caption{Final E787 data sample from 1995--97 after all cuts. One
clean \kpnn event is seen in the box. The remaining events are \kpp
background.
\label{fig:e787}}
\end{figure}

In E787's analysis of the 1995--97 data sample, one clean \kpnn event
lies in the signal box (see Fig.~\ref{fig:e787}),
with a measured  background of $0.08\pm0.02$ events.  Based on this
one event\cite{e787_pnn} the branching ratio is B(\kpnn) =
$1.5^{+3.4}_{-1.2} \times 10^{-10}$. From this measurement, a limit
 of $0.002 < |\vtd| < 0.04$ is determined; in addition, the following
limits on $\lambda_t \equiv V^*_{ts}\vtd$ can be set: 
$| Im \lambda_t | < 1.22\times10^{-3}$, 
$-1.10\times10^{-3} < Re \lambda_t  < 1.39\times10^{-3}$, and
$1.07\times10^{-4} < | \lambda_t | < 1.39\times10^{-3}$.
The final sensitivity of the E787 experiment, based on data
from 1995--98, should reach a factor of two further, to the
SM expectation for \kpnn.

A new experiment, E949, is under construction and will run in
2001--03. Taking advantage of the entire AGS proton flux and the
experience gained with the E787 detector, E949 with modest upgrades
should observe 10 SM events in a two year run. The background is
well-understood and is $\sim$10\% of the SM signal.

A proposal for a further factor of 10 improvement has been initiated
at FNAL. The CKM experiment (E905) plans to collect 100 SM events, with
B/S $\sim$ 10\%, in a two year run starting in $\sim$2005.  This
experiment will use a new technique, with K$^+$ decay-in-flight
and momentum/velocity spectrometers.

\boldmath
\subsection{\klpnn}
\unboldmath

The current best direct limit\cite{e799_pnn} on \klpnn comes from the
KTeV run in 1997: B(\klpnn) $< 5.9\times10^{-7}$ (90\% CL). 

An even more stringent limit can be derived in a model independent
way\cite{grossman} from the E787 measurement of \kpnn: 
\begin{equation}
B(\klpnn) < 4.4 \times  B(\kpnn) =  2.9\times10^{-9} \; \; (90\%\rm CL)
\label{eq:pnn}
\end{equation}

The next generation of \klpnn experiments will start with E391a at KEK,
which hopes to reach a sensitivity of $\sim10^{-10}$. While this does
not reach the SM level, the experiment will make a large step
forward and learn more about how to do this difficult experiment. The
plan would then be to move the detector to the JHF and push to a
sensitivity of ${\cal O} (10^{-13})$.  

Two other experiments propose to reach sensitivities of ${\cal O}
(10^{-13})$ --- E926 (KOPIO/RSVP) at BNL and E804 (KAMI) at FNAL. KAMI
plans to reuse the excellent CsI calorimeter from KTeV and to operate
at high kaon momentum to achieve good photon energy resolution and
efficiency.  The flux will increase substantially over KTeV due to
the Main Injector. KOPIO follows a different strategy.  The kaon
center of mass will be reconstructed using a bunched proton beam and a
very low momentum K$_L$ beam.  This gives 2 independent criteria to
reject background: photon veto and kinematics --- allowing background
levels to be directly measured from the data --- and gives further
confidence in the signal by measuring the spectra. The necessary flux will
by obtained using the entire AGS proton current. The low energy beam
also substantially reduces backgrounds from neutrons and other
sources. After three years of running, 65 SM events will be observed
with a S/B $\ge$ 2:1.

\section{Conclusions and Future Prospects}

The unprecedented sensitivities of the rare kaon decay experiments in
setting limits on LFV have constrained many extensions of the SM. The
discovery of \kpnn has opened the doors to measurements of the
unitarity triangle completely within the kaon system.  Significant
progress in the determination of the fundamental CKM parameters will
come from the generation of experiments that is starting now.
Comparison with the B-system will then over-constrain the triangle and
test the SM explanation of CP violation.

\section*{Acknowledgments}

I would like to thank members of several experiments for access to
their data and plots and for useful discussions, in particular, I
would like to thank Bill Molzon, Taku Yamanaka, Ron Ray, Bob
Tschirhart, Robin Appel, Hong Ma, Mike Zeller, Doug Bryman and Laurie
Littenberg.  This work was supported under U.S. Department of Energy
contract \#DE-AC02-98CH10886.

\end{document}